\documentclass[twocolumn,pra,showpacs,superscriptaddress,amssymb,amsmath,amsmath,floatfix]{revtex4-1}
\usepackage{graphicx}
\usepackage{epstopdf}
\usepackage{bm}
\usepackage{hyperref}
\usepackage{comment}
\usepackage{float}
\usepackage{cancel}

\hypersetup{%
   pdfpagemode=None, 
   pdfstartpage=1,
   pdfmenubar=true,
   pdftoolbar=true,
   colorlinks = true,
   linkcolor=blue,
   citecolor=blue,
   urlcolor=blue,
   bookmarksopen=false
 }

\newcommand{\be}{\begin{equation}}
\newcommand{\ee}{\end{equation}}

\begin{document}
\title{Highly polar molecules consisting of a copper or silver atom interacting with an alkali-metal or alkaline-earth-metal atom}

\author{Micha\l~\'Smia{\l}kowski}
\affiliation{Faculty of Physics, University of Warsaw, Pasteura 5, 02-093 Warsaw, Poland}
\affiliation{Faculty of Chemistry, University of Warsaw, Pasteura 1, 02-093 Warsaw, Poland}
\author{Micha\l~Tomza}
\email{michal.tomza@fuw.edu.pl}
\affiliation{Faculty of Physics, University of Warsaw, Pasteura 5, 02-093 Warsaw, Poland}

\date{\today}

\begin{abstract}

We theoretically investigate the properties of highly polar diatomic molecules containing $^2S$-state transition-metal atoms. We calculate potential energy curves, permanent electric dipole moments, spectroscopic constants, and leading long-range dispersion-interaction coefficients for molecules consisting of either a Cu or Ag atom interacting with an alkali-metal (Li, Na, K, Rb, Cs, Fr) or alkaline-earth-metal (Be, Mg, Ca, Sr, Ba, Ra) atom. We use \textit{ab initio} electronic structure methods, such as the coupled cluster and configuration interaction ones, with large Gaussian basis sets and small-core relativistic energy-consistent pseudopotentials. We predict that the studied molecules in the ground electronic state are strongly bound with highly polarized covalent or ionic bonds resulting in very large permanent electric dipole moments. We find that highly excited vibrational levels have maximal electric dipole moments, e.g., exceeding 13$\,$debye for CsAg and 6$\,$debye for BaAg. Results for Cu$_2$, Ag$_2$, and CuAg are also reported. The studied molecules may find application in ultracold dipolar many-body physics, controlled chemistry, or precision measurement experiments.

\end{abstract}

\maketitle

\section{Introduction}

Remarkable progress has been achieved in ultracold matter studies in the last decades. The advancement of cooling and trapping techniques has allowed reaching sub-millikelvin temperatures for gaseous ensembles of dozens of different kinds of atoms and simple molecules, finding application in fundamental research and emerging quantum technologies~\cite{GrossScience17,BohnScience17,DeMilleScience17}. Experiments with ultracold polar molecules reveal intriguing perspectives based on both complex internal molecular structure and intermolecular interactions~\cite{CarrNJP09,QuemenerCR12}. 
The rich internal structure can be employed in high-precision spectroscopic measurements to test fundamental physics, including searches for the electric dipole moment of the electron~\cite{AndreevNature18} and spatiotemporal variation of fundamental constants such as the electron-to-proton mass ratio~\cite{SchillerPRA05,DeMillePRL08a} and the fine structure constant~\cite{HudsonPRL06}, as well as tests of the quantum electrodynamics, parity violation, Lorentz symmetry, and general relativity~\cite{SafronovaRMP18}. Long-range and controllable intermolecular interactions between ultracold polar molecules allow studying ultracold chemistry, including quantum-controlled chemical reactions~\cite{OspelkausScience10,MirandaNatPhys11,TomzaPRL15}, and quantum many-body physics, including quantum simulation of quantum many-body Hamiltonians of increasing complexity~\cite{MicheliNatPhys06,BoNature13,DawidPRA18,AndereggS19}.

Ultracold molecules can be produced either directly by laser cooling~\cite{ShumanNature10}, buffer-gas~\cite{WeinsteinNat98} or sympathetic~\cite{SonNature20} cooling, Stark~\cite{MeerakkerCR12} or Zeeman~\cite{NareviciusCR12} deceleration, or velocity filtering~\cite{RangwalaPRA03} from higher temperatures, or indirectly by associating from ultracold atoms employing photoassociation~\cite{JonesRMP06} or magnetoassociation~\cite{KohlerRMP06}. Atomic species selected for pre-cooling and subsequent molecule formation have mostly been either alkali or alkaline-earth metals due to their electronic structure favorable for laser cooling. However, atoms of other elements have also been successfully laser-cooled. Bose-Einstein condensates of highly magnetic lanthanide Dy~\cite{LevPRL11} and Er~\cite{FerlainoPRL12} atoms, and transition-metal Cr~\cite{PfauPRL05} atoms were obtained at ultralow temperatures and employed in ground-breaking experiment~\cite{LahayeNature07,BaierScience16,SchmittNature16}. Magneto-optical cooling and trapping of other highly magnetic atoms such as Eu~\cite{DoylePRL97,InouePRA18}, Tm~\cite{SukachevPRA10}, and Ho~\cite{MiaoPRA14} were also realized. On the other hand, alkali-metal-like transition-metal Cu and Ag atoms were produced and trapped at ultralow temperatures using buffer-gas cooling and magnetic trapping~\cite{BrahmsPRL08} or magneto-optical cooling and trapping~\cite{UhlenbergPRA00}. All those atoms are available for the formation of new ultracold molecules with desirable properties. However, only the magnetoassociation into ultracold Er$_2$ dimers~\cite{FrischPRL15} and photoassociation into spin-polarized Cr$_2$ dimers~\cite{RuhrigPRA16} were experimentally demonstrated, while several heteronuclear paramagnetic and polar molecules formed of atoms with large magnetic dipole moments, such as CrRb~\cite{SadeghpourPRA10}, CrSr and CrYb~\cite{TomzaPRA13a}, ErLi~\cite{GonzalezPRA15},  EuK, EuRb, and EuCs~\cite{TomzaPRA14,ZarembaPRA18}, ErYb~\cite{KosickiNJP20}, and DyYb~\cite{FryePRX20} were theoretically proposed and studied.

Here, we propose the formation of ultracold highly polar diatomic molecules containing a transition-metal copper or silver atom interacting with an alkali-metal or alkaline-earth-metal atom. While such molecules have the ground-state electronic structure similar to alkali-metal or alkali-metal--alkaline-earth-metal dimers, they have a richer structure of excited electronic states owing to the possibility of $d$-electron excitations. A greater variety of excited electronic states may be beneficial for precision measurements~\cite{SafronovaRMP18}. Already, atomic clocks based on metastable states of Cu, Ag, and Au atoms were proposed for timekeeping and searching for new physics~\cite{Dzuba2020}, and the $^2S_{1/2}\to^2D_{5/2}$  clock transition in Ag was observed by two-photon laser spectroscopy~\cite{BadrPRA06}. RaCu and RaAg molecules were also proposed for measuring the electric dipole moment of the electron and the scalar-pseudoscalar interaction~\cite{SunagaPRA19}. Additionally, Cu and Ag atoms have high electronegativity as compared with alkali-metal and alkaline-earth-metal atoms, promising strong bonding and large permanent electric dipole moments of the considered molecules. While the interactions of Cu and Ag atoms with noble gases have been the subject of several experimental~\cite{JouvetJCP91,BrockCPL95,BrockJCP95,BrahmsPRL10} and theoretical~\cite{ZhangPRA08,TscherbulPRL11,LoreauJCP13} studies and the structure of the Cu$_2$, CuAg and Ag$_2$ dimers have been actively explored~\cite{BrownJMS78,BausJCP83,StollJCP84,HayJCP85,RohlfingJCP86,PageJCP91,SimardCPL91,BisheaJCP91,KramerCPL92,BeutelJCP93,PouJCP94,WangJMS02}, the interactions of Cu and Ag atoms with alkali-metal and alkaline-earth-metal metal atoms (and corresponding diatomic molecules) have been investigated in spectroscopic experiments occasionally~\cite{NeubertJCSF74,YehCPL93,PilgrimCPL95,StangassingerCPL97,BrockJCP97,RussonJCP97} and in theoretical calculations rarely~\cite{BeckmannCPL85,BauschlicherJCP87,LawsonJPC96,TongJPCA02,SunagaPRA19}.

In this paper, to fill this gap and to extend the range of species available for ultracold studies, we theoretically investigate the ground-state properties of highly polar diatomic molecules consisting of either a Cu or Ag atom interacting with an alkali-metal (Li, Na, K, Rb, Cs, Fr) or alkaline-earth-metal (Be, Mg, Ca, Sr, Ba, Ra) atom. We employ state-of-the-art \textit{ab initio} electronic structure methods, such as the coupled cluster and configuration interaction ones, with large Gaussian basis sets and small-core relativistic energy-consistent pseudopotentials to account for the scalar relativistic effects. We calculate potential energy curves, permanent electric dipole moments, spectroscopic constants, and leading long-range dispersion-interaction coefficients. We predict that the studied molecules in the ground electronic state are strongly bound with highly polarized covalent or ionic bonds resulting in significant permanent electric dipole moments, significantly larger than in alkali-metal molecules. We find that maximal electric dipole moments, exceeding 13$\,$debye for CsAg and 6$\,$debye for BaAg, are for highly excited vibrational levels. Results for Cu$_2$, Ag$_2$, and CuAg are also reported. We show that most of the investigated molecules in the ground state are stable against atom-exchange chemical reactions. Finally, we indicate their possible application in ultracold dipolar many-body physics, controlled chemistry, or precision measurement experiments.

The structure of the paper is the following. In Section~\ref{sec:theory}, we describe the employed computational methods. In Section~\ref{sec:results}, we present and discuss the obtained results. In section~\ref{sec:summary}, we provide a summary and outlook.

\section{Computational details}
\label{sec:theory}

The interaction of an open-shell copper or silver atom in the ground doublet $^2S$ electronic state with an open-shell alkali-metal atom, \textit{AM}, also in the lowest $^2S$ state, results in the ground molecular electronic state of the singlet $X^1\Sigma^+$ symmetry and the first excited electronic state of the triplet $a^3\Sigma^+$ symmetry of a \textit{AM}Cu or \textit{AM}Ag molecule. The interaction of a copper or silver atom in the $^2S$ electronic state with a closed-shell alkaline-earth-metal atom, \textit{AEM}, in the lowest singlet $^1S$ state, results in the ground molecular electronic state of the doublet $X^2\Sigma^+$ symmetry of a \textit{AEM}Cu or \textit{AEM}Ag molecule.

To calculate potential energy curves in the Born-Oppenheimer approximation, we adopt the computational scheme successfully applied to the ground electronic states of other diatomic molecules containing alkali-metal or alkaline-earth-metal atoms~\cite{TomzaPRA13a,TomzaPRA13b,TomzaPRA14,GronowskiPRA20,SmialkowskiPRA20}. The considered doublet $X^2\Sigma^+$ and triplet $a^3\Sigma^+$ molecular electronic states are well described at all internuclear distances by single-reference methods. Therefore, we compute them with the spin-restricted open-shell coupled cluster method restricted to single, double, and non-iterative triple excitations (RCCSD(T))~\cite{PurvisJCP82,KnowlesJCP93}. On the other hand, the singlet $X^1\Sigma^+$ molecular electronic states of the \textit{AM}Cu and \textit{AM}Ag molecules have single-reference nature at smaller internuclear distances and multireference nature at larger distances, which originates from the open-shell character of the interacting atoms. Therefore, we compute these electronic states with the RCCSD(T) method in the vicinity of the interaction potential well at short and intermediate distances and smoothly merge them with results obtained with the multireference configuration interaction method restricted to single and double excitations (MRCISD)~\cite{KnowlesTCA92} at larger distances. The MRCISD results are shifted to impose correct asymptotic energies. We use the switching function from Ref.~\cite{JanssenJCP09} over a distance of 2 bohr centered around 9-12 bohr depending on the system.

The interaction energy,~$E_\text{int}(R)$, at the internuclear distance $R$, is computed with the supermolecular method with the basis set superposition error corrected by using the Boys-Bernardi counterpoise correction~\cite{BoysMP70}
\begin{equation}
E_\text{int}(R)=E_{AB}(R)-E_{A}(R)-E_{B}(R)\,,
\label{eq:intenergy}
\end{equation}
where $E_{AB}(R)$ is the total energy of the molecule $AB$, and $E_{A}(R)$ and $E_{B}(R)$ are the total energies of the atoms $A$ and $B$ computed in the diatom basis set, all at the distance $R$.

The Li, Be, Na, and Mg atoms are described with the augmented correlation-consistent polarized weighted core-valence quintuple-$\zeta$ quality basis sets (aug-cc-pwCV5Z)~\cite{PrascherTCA10}. The scalar relativistic effects in heavier atoms are included by employing the small-core relativistic energy-consistent pseudopotentials (ECP) to replace the inner-shell electrons~\cite{DolgCR12}. The pseudopotentials from the Stuttgart library are used in all calculations. The Cu, Ag, K, Ca, Rb, Sr, Cs, Ba, Fr, and Ra atoms are described with the ECP10MDF, ECP28MDF, ECP10MDF, ECP10MDF, ECP28MDF, ECP28MDF, ECP46MDF, ECP46MDF, ECP78MDF, and ECP78MDF pseudopotentials~\cite{LimJCP06,DolgTCA98}, respectively, together with the aug-cc-pwCV5Z basis sets designed for those ECPs~\cite{HillJCP17,PetersonTCA05}. The atomic basis sets are additionally augmented in all calculations by the set of the $[3s3p2d2f1g]$ bond functions~\footnote{Bond function exponents, $s$: 0.9, 0.3, 0.1, $p$: 0.9, 0.3, 0.1, $d$: 0.6, 0.2, $f$: 0.6, 0.2, $g$: 0.3.} to accelerate the convergence towards the complete basis set limit~\cite{midbond}. The electrons of two outermost shells are correlated, i.e.~$3s^23p^63d^{10}4s^1$ from Cu, $4s^24p^64d^{10}5s^1$ from Ag, $(n-1)s^2(n-1)p^6{n}s^1$ from alkali-metal and  $(n-1)s^2(n-1)p^6{n}s^2$ from alkaline-earth-metal atoms.

The interaction potential between two neutral atoms in the electronic ground state is asymptotically dominated by the dispersion interaction of the form~\cite{JeziorskiCR94}
\begin{equation}\label{eq:E6}
E_\text{int}(R)=-\frac{C_6}{R^6}+\dots\,,
\end{equation}
where the leading $C_6$ coefficient is given by
\begin{equation}\label{eq:C6}
C_6=\frac{3}{\pi}\int_0^\infty \alpha_A(i\omega)\alpha_B(i\omega)d\omega\,,
\end{equation}
where $\alpha_{A(B)}(i\omega)$ is the dynamic electric dipole polarizbility of the $A(B)$ atom at the imaginary frequency $i\omega$. The dynamic polarizabilities at the imaginary frequency of the  alkali-metal and alkaline-earth-metal atoms are taken from Ref.~\cite{DerevienkoADNDT10}, whereas the dynamic polarizabilities of the Cu and Ag atoms are constructed as a sum over states using experimental energies~\cite{nist} and transition dipole moments from Refs.~\cite{TopcuPRA06,ZhangPRA08}.

The permanent electric dipole moments and static electric dipole and quadrupole polarizabilities are calculated with the finite field approach using the RCCSD(T) method (for the $X^1\Sigma^+$ states of the $AM$Cu and $AM$Ag using the RCCSD(T) and MRCISD methods and merged similarly as potential energy curves). The $z$ axis is chosen along the internuclear axis, oriented from the Cu or Ag atom to the alkali-metal or alkaline-earth-metal atom. The vibrationally averaged dipole moments are calculated as expectation values of $R$-dependent dipole moment functions with radial vibrational wavefunction.

All electronic structure calculations are performed with the \textsc{Molpro} package of \textit{ab initio} programs~\cite{Molpro,MOLPRO-WIREs}. Vibrational eigenenergies and eigenstates are calculated using numerically exact diagonalization of the Hamiltonian for the nuclear motion within the discrete variable representation (DVR) on the non-equidistant grid~\cite{TiesingaPRA98}. Atomic masses of the most abundant isotopes are assumed.

\begin{table}[tb!]
\caption{Characteristics of alkali-metal and alkaline-earth-metal atoms: the static electric dipole polarizability $\alpha$, the static electric quadrupole polarizability $\beta$, the ionization potential IP, and the lowest $S$--$P$ excitation energy (${}^2S$--${}^2P$ for alkali-metal, Cu, and Ag atoms and ${}^1S$--${}^3P$ for alkaline-earth-metal atoms). Present theoretical values are compared with the most accurate available experimental or theoretical data. Experimental excitation energies are averaged on spin--orbit manifolds. \label{tab:atoms}}  
\begin{ruledtabular}
\begin{tabular}{lllll}
Atom & $\alpha\,$(a.u.) & $\beta\,$(a.u.) & IP$\,$(cm$^{-1}$) & $S$--$P$$\,$(cm$^{-1}$)\\
\hline
Li &  164.2 &  1418 & 43628 &  14902 \\
   &  164.2~\cite{MiffreEPJ06} & 1423~\cite{YanPRA96} & 43487~\cite{nist} & 14904~\cite{nist} \\
Na &  163.9 &  1881 & 41384 & 16769 \\
   &  162.7~\cite{EkstromPRA95} & 1895~\cite{KaurPRA15} & 41449~\cite{nist} &  16968~\cite{nist}     \\
K  &  289.6 &  4962 & 35153  & 12876 \\
   &  290.0~\cite{GregoirePRA15}  & 4947~\cite{KaurPRA15} & 35010~\cite{nist} & 13024~\cite{nist}      \\
Rb &  317.4 &  6485 & 33649  & 12573 \\
   &  320.1~\cite{GregoirePRA15} &  6491~\cite{KaurPRA15} & 33691~\cite{nist} &  12737~\cite{nist}     \\
Cs &  391.1 & 10498 & 31428 & 11318 \\
   &  401.2~\cite{GregoirePRA15}   & 10470~\cite{PorsevJCP03} & 31406~\cite{nist} &  11548~\cite{nist}     \\
Fr &  325.8 & 9225 & 32428 & 12452 \\
   &  317.8~\cite{dereviankoPRL99} & - & 32849~\cite{nist} & 12237~\cite{nist} \\
Be &  37.7 & 300 & 75169 & 43281 \\
   & 37.7~\cite{mitroyPRA03} & 301~\cite{PorsevJETP} & 75193~\cite{nist} & 42565~\cite{nist} \\
Mg &   71.5 &   816 & 61569  & 21967 \\
   &   71.3~\cite{PorsevJETP}  &  812~\cite{PorsevJETP} & 61671~\cite{nist} & 21891~\cite{nist}     \\
Ca &  156.2 &  3003 & 49378  & 15092 \\
   &  157.1~\cite{PorsevJETP}  &  3081~\cite{PorsevJETP} & 49306~\cite{nist}  & 15263~\cite{nist}      \\
Sr &  198.6 &  4576 & 45876  & 14518 \\
   &  197.2~\cite{PorsevJETP} &  4630~\cite{PorsevJETP} & 45932~\cite{nist} & 14705~\cite{nist}   \\
Ba &  274.3 &  8628 & 41915 & 12891 \\
   &  273.5~\cite{PorsevJETP} &  8900~\cite{PorsevJETP} & 42035~\cite{nist} & 13083~\cite{nist} \\
Ra &  250.5 &  7480 & 42208 & 14647 \\
   & 248.6~\cite{LimPRA04} & 7147~\cite{TeodoroJCP15} & 42573~\cite{nist} & 15391~\cite{nist} \\
Cu &  45.9 &  325 & 62406 & 29792 \\
   & 46.5~\cite{neogradyIJQC97} & 332~\cite{ZhangPRA08} & 62317~\cite{nist} & 30691~\cite{nist} \\
Ag &  50.1 &  385 & 61249 & 28565 \\
   & 52.5~\cite{neogradyIJQC97} & 392~\cite{ZhangPRA08} & 61106~\cite{nist} & 30127~\cite{nist} 
\end{tabular}
\end{ruledtabular}
\end{table}

\section{Results and discussion}
\label{sec:results}

\subsection{Atomic properties}

An accurate description of atoms is essential for a proper evaluation of interatomic interactions. Therefore, to determine the ability of the employed \textit{ab initio} approaches to produce accurate results, we examine the electronic properties of investigated atoms, which also decide the long-range interaction coefficients crucial for ultracold physics and chemistry. 

Table~\ref{tab:atoms} collects the static electric dipole and quadrupole polarizabilities, ionization potentials, and the lowest $S$--$P$ excitation energies of the alkali-metal, alkaline-earth-metal, Cu, and Ag atoms. Present theoretical values are compared with the most accurate available theoretical or experimental data. The calculated static electric dipole and quadrupole polarizabilities coincide with previous data within 0-10.7$\,$a.u.~and 1-333$\,$a.u.~that correspond to an error of 0-4.6$\,\%$ (1.1\% on average) and 0.1-4.5$\,\%$ (1.4\% on average), respectively. The ionization potentials and the lowest $S$--$P$ excitation energies agree with experiential data within 22-421$\,$cm$^{-1}$ and 2-1562$\,$cm$^{-1}$ that is 0.1-1.3$\,\%$ (0.31$\,\%$ on average) and 0.01-5.2$\,\%$ (1.9\% on average), respectively. The description of the heaviest alkali-metal and alkaline-earth-metal atoms and the Cu and Ag atoms is the most challenging.

The overall agreement between calculated atomic properties and the most accurate available experimental or theoretical data is very good. It confirms that the employed electronic structure methods, basis sets, and energy-consistent pseudopotentials properly treat the relativistic effects and reproduce the correlation energy while being close to being converged in the size of the basis function set. Thus, the used methodology should also provide an accurate description of interatomic interactions and molecular properties investigated in the next subsections.  Based on the above, test calculations for smaller basis sets, and our previous experience~\cite{SmialkowskiPRA20}, we estimate the total uncertainty of the calculated ground-state $X^1\Sigma^+$ and $X^2\Sigma^+$ potential energy curves at the equilibrium distance to be of the order of 250-600$\,$cm$^{-1}$ that corresponds to 2-5\% of the interaction energy. The quality of employed energy-consistent pseudopotentials and basis sets, followed by the lack of the exact treatment of the triple and higher excitations in the employed CCSD(T) method, are primary limiting factors. The uncertainty of the long-range interaction coefficients is of the same order of magnitude. The relative uncertainty of the weakly bound $a^3\Sigma^+$ potential energy curves is expected to be a bit larger.

\begin{figure*}[tb]
\begin{center}
\includegraphics[width=\textwidth]{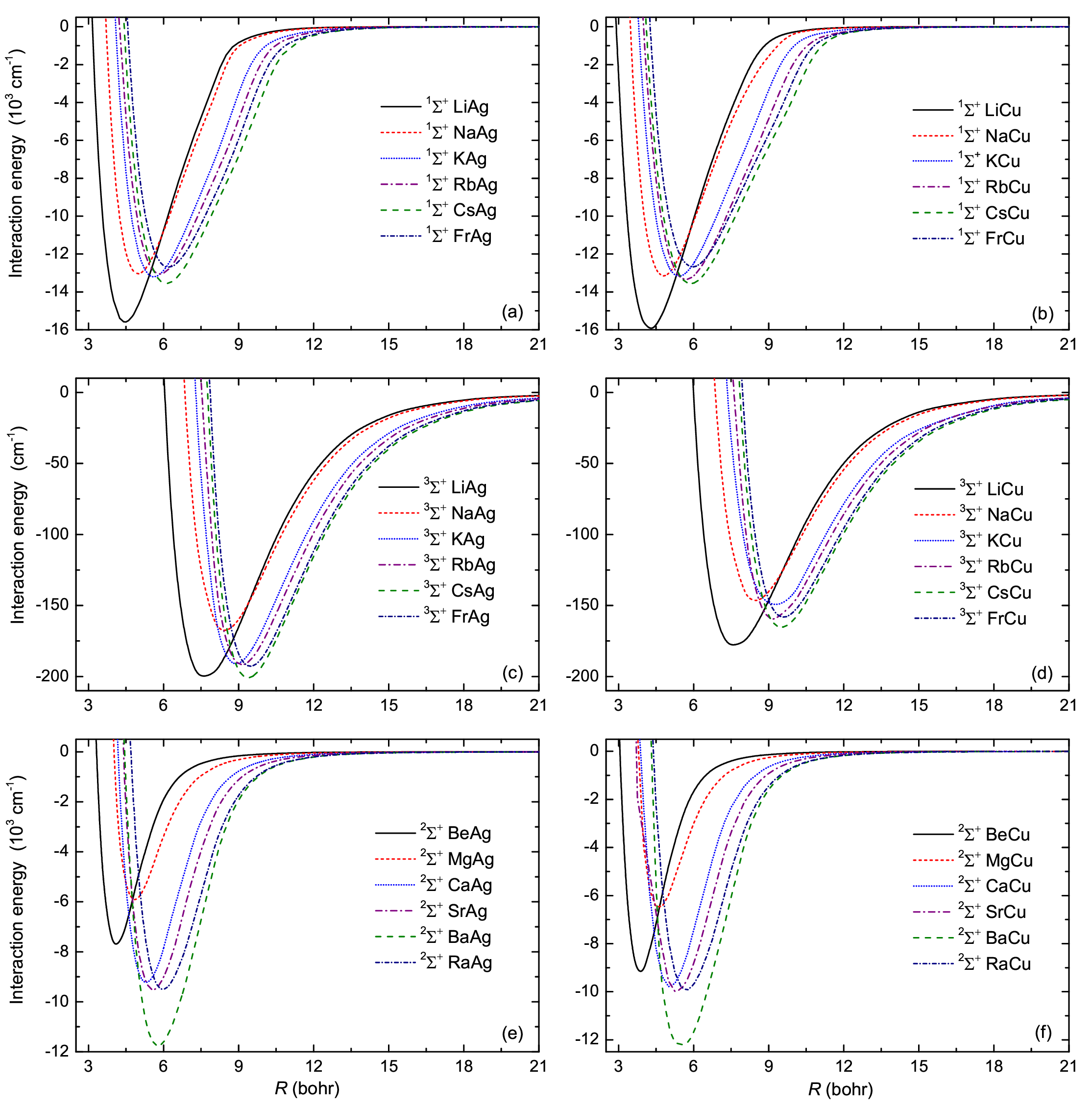}
\end{center}
\caption{Potential energy curves of the $X^1\Sigma^+$ electronic states of the \textit{AM}Ag~(a) and \textit{AM}Cu~(b) molecules, the $a^3\Sigma^+$ electronic states of the \textit{AM}Ag~(c) and \textit{AM}Cu~(d) molecules, and  the $X^2\Sigma^+$ electronic states of the \textit{AEM}Ag~(e) and \textit{AEM}Cu~(f) molecules.}
\label{fig:PES}
\end{figure*}

\subsection{Potential energy curves}

\begin{table*}[tb]
\caption{Characteristics of the \textit{AM}Ag and \textit{AM}Cu molecules in the $X^1\Sigma^+$ and $a^3\Sigma^+$ electronic states and the \textit{AEM}Ag and \textit{AEM}Cu molecules in the $X^2\Sigma^+$ electronic state: equilibrium interatomic distance~$R_e$, well depth~$D_e$, harmonic constant~$\omega_e$, rotational constant~$B_e$, permanent electric dipole moment~$d_e$, and fraction of its maximal possible value~$d_e/d_\text{max}$, parallel and perpendicular components of the static electric dipole polarizability~$\alpha_e^\parallel$ and $\alpha_e^\perp$, number of vibrational levels~$N_v$, and long-range dispersion-interaction coefficient~$C_6$.}
\label{tab:const} 
\begin{ruledtabular}
\begin{tabular}{lrrrrrrrrrr}
Molecule & $R_e\,$(bohr) & $D_e\,$(cm$^{-1}$) & $\omega_e\,$(cm$^{-1}$) & $B_e\,$(cm$^{-1}$) & $d_e\,$(D) & $d_e/d_\text{max}$ & $\alpha_e^\parallel$ (a.u.) & $\alpha_e^\perp\,$(a.u.) & $N_v$ & $C_6\,$(a.u.) \\
\hline
\multicolumn{9}{c}{$X^1\Sigma^+$ electronic state} \\
LiAg & 4.460 & 15592 & 389.7 & 0.4591 & 5.21 & 0.459 & 113.5 & 82.3 & 63 & 567 \\
NaAg & 4.985 & 13040 & 212.6 & 0.1276 & 5.98 & 0.472 & 148.7 & 99.3 & 96 & 611 \\
KAg & 5.609 & 13204 & 147.4 & 0.0667 & 8.50 & 0.597 & 161.8 & 106.2 & 133 & 937 \\
RbAg & 5.845 & 13058 & 109.7 & 0.0369 & 9.03 & 0.608 & 183.9 & 110.5 & 176 & 1031 \\
CsAg & 6.112 & 13562 & 92.3 & 0.0269 & 9.75 & 0.628 & 200.6 & 113.6 & 214 & 1234 \\
FrAg & 6.190 & 12700 & 84.2 & 0.0215 & 9.20 & 0.585 & 235.1 & 116.7 & 219 & 1116 \\ 
LiCu & 4.259 & 15959 & 399.1 & 0.5257 & 5.05 & 0.467 & 102.4 & 76.9 & 63 & 523 \\ 
NaCu & 4.796 & 13158 & 223.7 & 0.1555 & 5.78 & 0.474 & 121.3 & 91.9 & 94  & 564 \\
KCu & 5.410 & 13180 & 158.3 & 0.0855 & 8.24 & 0.599 & 142.6 & 100.6 & 124 & 864 \\
RbCu & 5.687 & 13355 & 123.6 & 0.0515 & 8.81 & 0.609 & 165.3 & 104.4 & 161 & 950 \\
CsCu & 5.874 & 13566 & 106.2 & 0.0408 & 9.29 & 0.622 & 191.7 & 107.7 & 187 & 1138 \\
FrCu & 5.958 & 12685 & 100.5 & 0.0345 & 8.80 & 0.581 & 222.9 & 109.8 & 186 & 1030 \\
Cu$_2$ & 4.337 & 16329 & 253.9 & 0.1017 & 0 & 0 & 124.4 & 61.7 & 117 & 221 \\
AgCu & 4.587 & 15048 & 234.0 & 0.0722 & 0.037 & 0.008 & 140.8 & 68.48 & 138 & 239\\
Ag$_2$ & 4.904 & 13902 & 188.2 & 0.0468 & 0 & 0 & 159.8 & 79.3 & 138 & 258 \\
\hline
\multicolumn{9}{c}{$a^3\Sigma^+$ electronic state} \\
LiAg & 7.649 & 202 & 34.7 & 0.1563 & 0.077 & 0.0040 & 334.2 & 185.9 & 12 & 567 \\
NaAg & 8.427 & 168 & 19.8 & 0.0448 & 0.126 & 0.0059 & 401.9 & 269.2 & 18 & 611 \\
KAg & 8.886 & 192 & 16.8 & 0.0267 & 0.149 & 0.0066 & 458.6 & 302.9 & 25 & 937 \\
RbAg & 9.121 & 192 & 13.0 & 0.0153 & 0.148 & 0.0064 & 517.5 & 357.9 & 32 & 1031 \\
CsAg & 9.362 & 202 & 11.6 & 0.0116 & 0.130 & 0.0055 & 584.5 & 399.7 & 38 & 1234 \\
FrAg & 9.451 & 193 & 10.6 & 0.0093 & 0.130 & 0.0054 & 612.8 & 342.8 & 41 & 1116 \\
LiCu & 7.634 & 179 & 32.8 & 0.1637 & 0.017 & 0.0009 & 303.9 & 172.2 & 11 & 523 \\
NaCu & 8.453 & 147 & 19.7 & 0.0500 & 0.071 & 0.0033 & 385.3 & 255.7 & 16 & 564 \\
KCu & 8.973 & 162 & 16.6 & 0.0311 & 0.067 & 0.0029 & 429.3 & 297.5 & 21 & 864 \\
RbCu & 9.228 & 161 & 13.5 & 0.0196 & 0.063 & 0.0027 & 487.9 & 351.4 & 26 & 950 \\
CsCu & 9.491 & 166 & 12.2 & 0.0156 & 0.037 & 0.0015 & 550.4 & 389.0 & 30 & 1138 \\
FrCu & 9.579 & 159 & 11.5 & 0.0134 & 0.049 & 0.0020 & 469.4 & 339.1 & 31 & 1030 \\
Cu$_2$ & 5.082 & 548 & 70.7 & 0.0741 & 0 & 0 & 176.8 & 83.1 & 29 & 221 \\
AgCu & 5.566 & 447 & 47.5 & 0.0490 & 0.028 & 0.0050 & 174.2 & 82.3 & 32 & 239 \\
Ag$_2$ & 5.937 & 459 & 38.0 & 0.0319 & 0 & 0 & 173.9 & 80.5 & 38 & 258 \\
\hline
\multicolumn{9}{c}{$X^2\Sigma^+$ electronic state} \\
BeAg & 4.109 & 7722 & 438.9 & 0.4290 & -0.71 & 0.068 & 131.7 & 64.1 & 37 & 231 \\
MgAg & 4.829 & 5995 & 230.0 & 0.1318 & 1.09 & 0.089 & 177.3 & 92.8 & 56 & 400 \\
CaAg & 5.292 & 9300 & 179.0 & 0.0739 & 2.62 & 0.195 & 213.6 & 163.4 & 94 & 737 \\
SrAg & 5.568 & 9586 & 132.1 & 0.0402 & 3.57 & 0.253 & 258.2 & 238.1 & 128 & 889 \\
BaAg & 5.769 & 11822 & 114.4 & 0.0300 & 4.52 & 0.309 & 290.6 & 292.3 & 169 & 1136 \\
RaAg & 5.959 & 9563 & 100.6 & 0.0234 & 5.08 & 0.336 & 293.9 & 297.3 & 164 & 1053\\
BeCu & 3.916 & 9108 & 505.9 & 0.4978 & -0.81 & 0.081 & 118.5 & 59.5 & 38 & 214 \\
MgCu & 4.606 & 6527 & 252.0  & 0.1634 & 0.94 & 0.080 & 157.4 & 84.3 & 54 & 371 \\
CaCu & 5.054 & 9796 & 197.2 & 0.0964 & 2.32 & 0.180 & 183.8 & 157.2 & 88 & 681 \\
SrCu & 5.328 & 10006 & 153.0 & 0.0578 & 3.25 & 0.240 & 226.6 & 236.1 & 115 & 821 \\
BaCu & 5.484 & 12437 & 137.1 & 0.0463 & 4.01 & 0.288 & 277.2 & 295.4 & 148 & 1049 \\
RaCu & 5.700 & 9946 & 121.6 & 0.0376 & 4.65 & 0.321 & 275.9 & 299.4 & 138 & 972 \\
\end{tabular}
\end{ruledtabular}
\end{table*}

The computed potential energy curves of the $X^1\Sigma^+$ symmetry for the \textit{AM}Ag and \textit{AM}Cu molecules, the $a^3\Sigma^+$ symmetry for the \textit{AM}Ag and \textit{AM}Cu molecules, and the $X^2\Sigma^+$ symmetry for the \textit{AEM}Ag and \textit{AEM}Cu molecules are presented in Fig.~\ref{fig:PES}. Calculations are performed for all alkali-metal (\textit{AM}=Li, Na, K, Rb, Cs, Fr) and alkaline-earth-metal (\textit{AEM}=Be, Mg, Ca, Sr, Ba, Ra) atoms. The corresponding long-range dispersion-interaction $C_6$ coefficients and spectroscopic characteristics such as the equilibrium interatomic distance~$R_e$, well depth~$D_e$, harmonic constant~$\omega_e$, rotational constant~$B_e$, and number of vibrational levels~$N_v$ (for $j=0$) are collected in Table~\ref{tab:const}. 

All potential energy curves presented in Fig.~\ref{fig:PES} show a smooth behavior with well-defined minima. Surprisingly, the potential energy curves for the \textit{AM}Ag and \textit{AM}Cu molecules exhibit very similar shapes with similar equilibrium interatomic distances and well depths, which do not depend significantly on involved alkali-metal atoms (contrary to properties of alkali-metal dimers~\cite{AymarJCP05,DeiglmayrJCP08,ZuchowskiPRA10,TomzaPRA13b}). This suggests that the nature of their chemical bonds is mostly determined by the properties of the Ag and Cu atoms. 

The \textit{AM}Ag and \textit{AM}Cu molecules in the ground $X^1\Sigma^+$ electronic state are the most strongly bound with the well depths between 12700$\,$cm$^{-1}$ for FrAg and 15592$\,$cm$^{-1}$ for LiAg among the \textit{AM}Ag molecules, and between 12685$\,$cm$^{-1}$ for FrCu and 15959$\,$cm$^{-1}$ for LiCu among the \textit{AM}Cu molecules. The LiAg and LiCu molecules clearly exhibit the strongest chemical bonds in both groups, with well depths over 20$\,$\% larger than their analogs, which in turn do not differ by more than 5$\,$\%. Their equilibrium distances systematically increase with increasing the atomic number of the alkali-metal atom and take values between 4.46$\,$bohr for LiAg and 6.19$\,$bohr for FrAg among the \textit{AM}Ag molecules and between 4.26$\,$bohr for LiCu and 5.96$\,$bohr for FrCu among the \textit{AM}Cu molecules. The number of vibrational levels is between 63 for LiCu or LiAg and 219 for FrAg among the molecules in the $X^1\Sigma^+$ state.

The \textit{AM}Ag and \textit{AM}Cu molecules in the first-excited $a^3\Sigma^+$ electronic state are weakly bound van der Waals complexes with the well depths between 168$\,$cm$^{-1}$ for NaAg and 202$\,$cm$^{-1}$ for LiAg and CuAg among the \textit{AM}Ag molecules, and between 147$\,$cm$^{-1}$ for NaCu and 179$\,$cm$^{-1}$ for LiCu among the \textit{AM}Cu molecules. Their equilibrium distances systematically increase with increasing the atomic number of the alkali-metal atom and take values between 7.65$\,$bohr for LiAg and 9.45$\,$bohr for FrAg among the \textit{AM}Ag molecules and between 7.63$\,$bohr for LiCu and 9.58$\,$bohr for FrCu among the \textit{AM}Cu molecules. The number of vibrational levels is between 11 for LiCu and 41 for FrAg among the molecules in the $a^3\Sigma^+$ state.

The \textit{AM}Ag and \textit{AM}Cu molecules in the ground $X^1\Sigma^+$ electronic state are significantly more strongly bound than analogous alkali-metal molecules~\cite{ZuchowskiPRA10}, while the \textit{AM}Ag and \textit{AM}Cu molecules in the ground $a^3\Sigma^+$ electronic state are slightly less bound than analogous alkali-metal molecules~\cite{TomzaPRA13b}. All studied \textit{AM}Ag and \textit{AM}Cu molecules have shorter equilibrium distances than the corresponding homo- or heteronuclear alkali-metal dimers in respective electronic states.

The \textit{AEM}Ag and \textit{AEM}Cu molecules in the ground $X^2\Sigma^+$ electronic state are strongly bound with the well depths between 5995$\,$cm$^{-1}$ for MgAg and 11822$\,$cm$^{-1}$ for BaAg among the \textit{AEM}Ag molecules, and between 6527$\,$cm$^{-1}$ for MgCu and 12437$\,$cm$^{-1}$ for BaCu among the \textit{AEM}Cu molecules. The potential energy curves for the \textit{AEM}Ag and \textit{AEM}Cu molecules present a greater variety of well depths than the \textit{AM}Ag and \textit{AM}Cu ones. Their equilibrium distances, shorter than for other molecules, systematically increase with increasing the atomic number of the alkaline-earth-metal atom and take values between 4.11$\,$bohr for BeAg and 5.96$\,$bohr for RaAg among the \textit{AEM}Ag molecules and between 3.92$\,$bohr for BeCu and 5.70$\,$bohr for RaCu among the \textit{AEM}Cu molecules. The number of vibrational levels is between 34 for BeAg and 164 for BaAg among the molecules in the $X^2\Sigma^+$ state. The \textit{AEM}Ag and \textit{AEM}Cu molecules in the ground $X^2\Sigma^+$ electronic state are significantly more strongly bound with shorter equilibrium distances than analogous alkali-metal--alkaline-earth-metal molecules~\cite{PototschnigPCCP16}. 

The calculated long-range dispersion-interaction $C_6$ coefficients are smaller for investigated molecules than for analogous alkali-metal and alkaline-earth-metal molecules~\cite{PorsevJETP} because the polarizabilities of the Ag and Cu atoms are a few times smaller than the polarizabilities of alkali-metal and alkaline-earth-metal atoms (cf.~Table~\ref{tab:atoms}).

Among the investigated molecules, only a few have already been studied experimentally using photoionization spectroscopy~\cite{NeubertJCSF74,YehCPL93,PilgrimCPL95,StangassingerCPL97,BrockJCP97}. For LiAg, the equilibrium distance of 4.55$\,$bohr, the potential well depth of 15413(30)$\,$cm$^{-1}$, and the harmonic constant of 389.0$\,$cm$^{-1}$ were measured~\cite{PilgrimCPL95,BrockJCP97} in good agreement with the present values of 4.46$\,$bohr, 15592$\,$cm$^{-1}$, and 389.7$\,$cm$^{-1}$, respectively. For LiCu, the equilibrium distance of 4.27$\,$bohr, the potential well depth of 15961(12)$\,$cm$^{-1}$, and the harmonic constant of 465.9$\,$cm$^{-1}$ were measured~\cite{BrockJCP97,RussonJCP97} in good agreement with the present values of 4.26$\,$bohr, 15959$\,$cm$^{-1}$, and 399.1$\,$cm$^{-1}$, respectively. For NaAg, the potential well depth of at least 12932$\,$cm$^{-1}$ and the harmonic constant of 210$\,$cm$^{-1}$ were measured~\cite{StangassingerCPL97} in good agreement with the present values of 13040$\,$cm$^{-1}$ and 212.6$\,$cm$^{-1}$, respectively. The overall agreement with the spectroscopic studies confirms that similar high accuracy of present calculations may be expected for other molecules. The present theoretical results agree much better with the experimental measurements than previous calculations~\cite{BeckmannCPL85,BauschlicherJCP87,LawsonJPC96,TongJPCA02}, which underestimated well depths and overestimated equilibrium distances because they employed smaller basis sets and lower-level methods.   

The large binding energies and short equilibrium distances of the investigated molecules in their ground electronic states indicate the highly polarized covalent or even ionic nature of their chemical bonds~\cite{PilgrimCPL95,StangassingerCPL97,BrockJCP97} and significant stabilizing contribution of the electrostatic and induction interactions. The large difference of the electronegativity of the Ag or Cu atoms and the alkali-metal or alkaline-earth-metal atoms is responsible for a significant bond polarization and considerable contribution of the \textit{AM}$^+$Ag$^-$, \textit{AEM}$^+$Ag$^-$, \textit{AM}$^+$Cu$^-$, and \textit{AEM}$^+$Cu$^-$ ionic configurations to their ground state bonds~\cite{Pauling1960}. The electronegativity by Pauling scale~\cite{Pauling1960} of the Ag (1.93) and Cu (1.9) atoms is typically twice larger than that of the alkali-metal (0.79-0.98) and alkaline-earth-metal (0.89-1.57) atoms. This large difference, which is significantly larger than the variation of alkali-metal atoms' electronegativity, is responsible for the similarity of potential energy curves observed in Fig.~\ref{fig:PES}. Phenomenological models based on the difference of the electronegativities imply the ionic character of about 20-30$\,\%$ for the investigated molecules, except ones involving the lightest alkaline-earth-metal atoms~\cite{Pauling1960}. A considerable ionic contribution to the ground state bonding is consistent with a relatively small energy separation between the ion-pair asymptote and the asymptote of neutral ground state atoms in the studied molecules. This energy separation is given by the difference of the relatively low ionization potential of the alkali-metal or alkaline-earth-meta atoms and the high electron affinity of the Ag or Cu atoms. However, the multireference calculations for excited states show that the calculated electronic states are well separated (by at least 6000$\,$cm$^{-1}$) from excited electronic states. Additionally, our comparative multireference configuration interaction and higher-level coupled cluster calculations (following the approach presented for NaLi in Ref.~\cite{GronowskiPRA20}) confirm that all the studied electronic states are well described by the single-reference methods in the vicinity of the interaction potential well, and inclusion of higher-level excitation in the coupled cluster method is not necessary.  The nature of the chemical bonds is further analyzed in the following subsection.

\begin{figure}[tb]
\begin{center}
\includegraphics[width=\columnwidth]{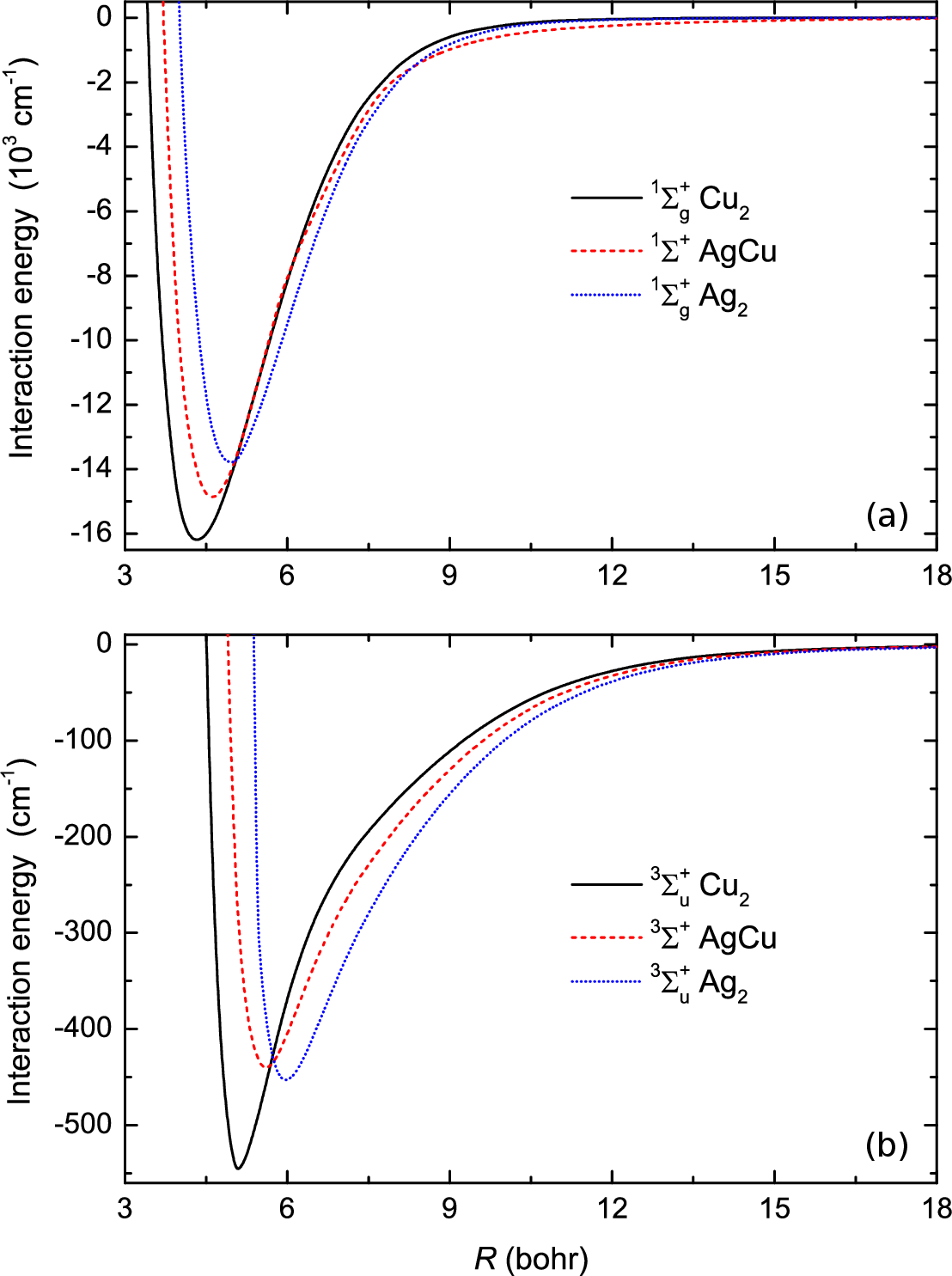}
\end{center}
\caption{Potential energy curves of the lowest (a) singlet $X^1\Sigma^+_g$/$X^1\Sigma^+$ and (b) triplet $a^3\Sigma^+_g$/$a^3\Sigma^+$ electronic states of the Cu$_2$, Ag$_2$, and AgCu molecules.}
\label{fig:AgCu}
\end{figure}

For the completeness of the analysis, we also calculate the properties of the Cu$_2$, Ag$_2$, and AgCu molecules in their ground $X^1\Sigma^+_g$ ($X^1\Sigma^+$ for AgCu) and lowest-excited $a^3\Sigma^+_u$ ($a^3\Sigma^+$ for AgCu) electronic states. Corresponding potential energy curves are presented in Fig.~\ref{fig:AgCu} and spectroscopic characteristics are collected in Table~\ref{tab:const}. The Cu$_2$, Ag$_2$, and AgCu dimers exhibit short, strong molecular bonding in the $X^1\Sigma^+$ state and weak van der Waals bonding in the a$^3\Sigma^+$ state, similarly to the \textit{AM}Ag and \textit{AM}Cu molecules. However, while the electrostatic and induction interactions dominantly stabilize the \textit{AM}Ag and \textit{AM}Cu molecules in the $X^1\Sigma^+$ state, the correlation of electrons from the $d$ orbitals of Cu and Ag atoms stabilizes the Cu$_2$, Ag$_2$ and AgCu molecules in the ground state~\cite{BausJCP83}. 

\begin{figure*}[tb]
\begin{center}
\includegraphics[width=\textwidth]{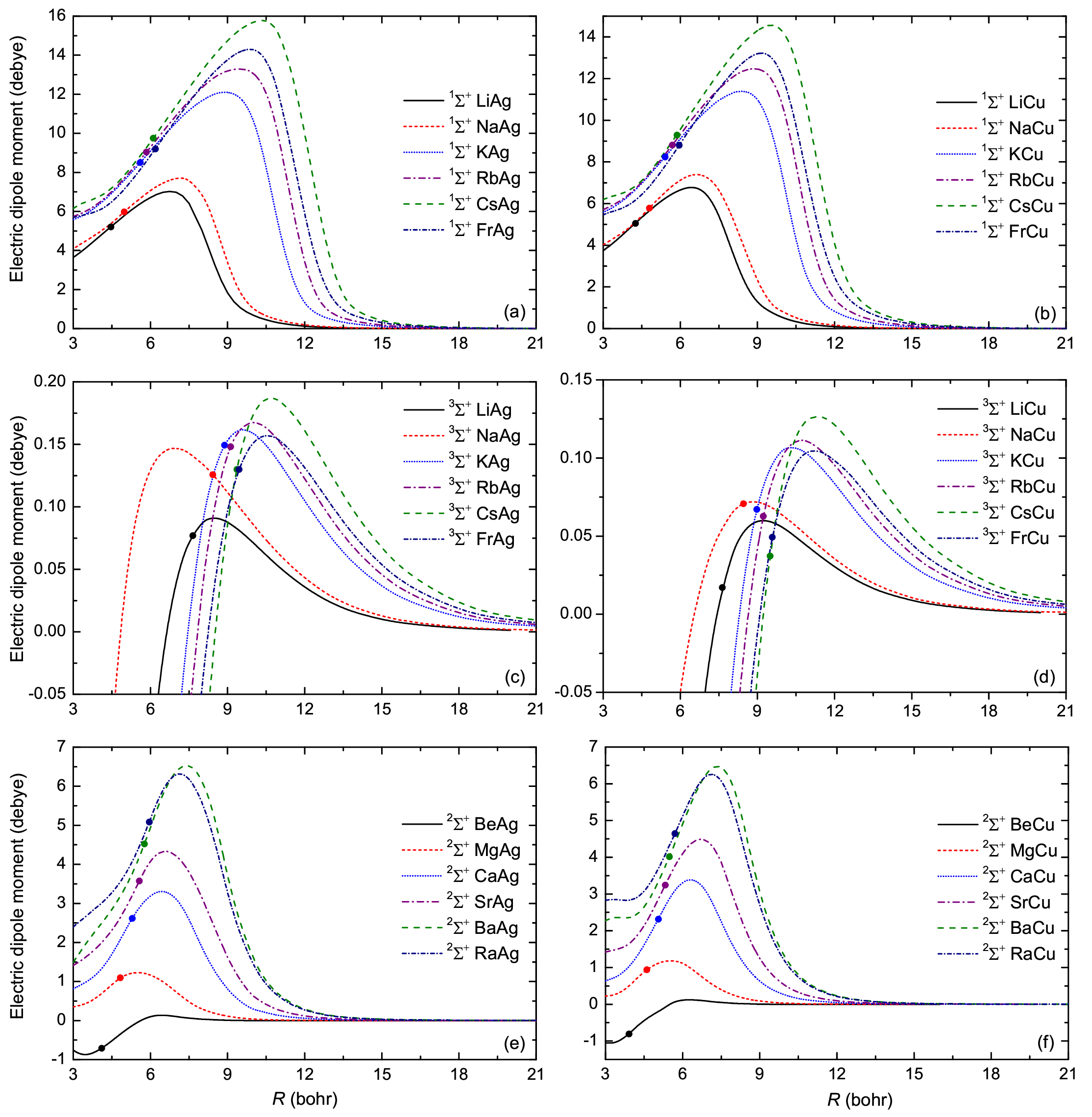}
\end{center}
\caption{Permanent electric dipole moments of the \textit{AM}Ag~(a) and \textit{AM}Cu~(b) molecules in the $X^1\Sigma^+$ electronic states, the \textit{AM}Ag~(c) and \textit{AM}Cu~(d) molecules in the $a^3\Sigma^+$ electronic states, and the \textit{AEM}Ag~(d) and \textit{AEM}Cu~(e) molecules in the $X^2\Sigma^+$ electronic states. The points indicate values for equilibrium distances.}
\label{fig:DP}
\end{figure*}

The calculated well depths of 16329$\,$cm$^{-1}$, 13902$\,$cm$^{-1}$, and 15048$\,$cm$^{-1}$, for Cu$_2$, Ag$_2$, and AgCu, agree well with experimental measurements of 16760(200)$\,$cm$^{-1}$~\cite{RohlfingJCP86}, 13403(250)$\,$cm$^{-1}$~\cite{BeutelJCP93}, and 14149(800)$\,$cm$^{-1}$~\cite{BisheaJCP91}, respectively. Similarly, the calculated harmonic constants of 253.9$\,$cm$^{-1}$, 188.2$\,$cm$^{-1}$, and 234.0$\,$cm$^{-1}$, for Cu$_2$, Ag$_2$, and AgCu, agree well with experimental values of 266.4(6)$\,$cm$^{-1}$~\cite{RohlfingJCP86}, 192.4$\,$cm$^{-1}$~\cite{BrownJMS78}, and 229.2(3)$\,$cm$^{-1}$~\cite{BisheaJCP91}, respectively. Such a good agreement additionally validates the accuracy of the present results, which, also in the case of the dimers of noble-metal atoms, are much more accurate than older calculations~\cite{StollJCP84,HayJCP85,PouJCP94,WangJMS02} and agree well with previous accurate results \cite{PetersonTCA05}.

\subsection{Permanent electric dipole moments}

\begin{figure}[tb]
\begin{center}
\includegraphics[width=\columnwidth]{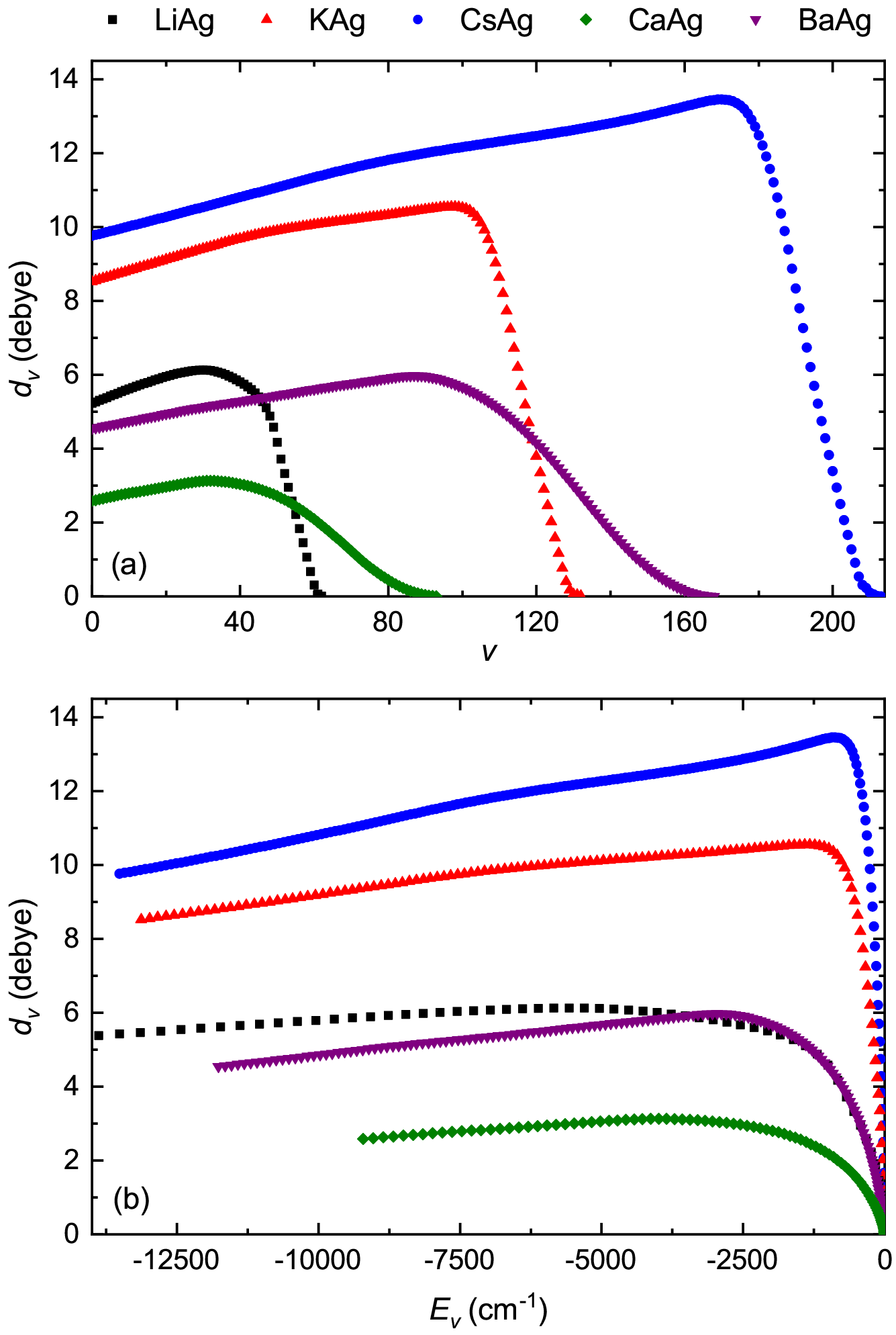}
\end{center}
\caption{Permanent electric dipole moment of the LiAg, KAg, CsAg, CaAg, and BaAg molecules in different vibrational levels of their ground electronic state as a function of the vibrational quantum number (a) and binding energy (b).}
\label{fig:dv}
\end{figure}

Permanent electric dipole moments as functions of the interatomic distance for the \textit{AM}Ag and \textit{AM}Cu molecules in the $X^1\Sigma^+$ electronic states, the \textit{AM}Ag and \textit{AM}Cu molecules in the $a^3\Sigma^+$ electronic states, and the \textit{AEM}Ag\ and \textit{AEM}Cu molecules in the $X^2\Sigma^+$ electronic states are presented in Fig.~\ref{fig:DP}. The corresponding values for equilibrium distances are collected in Table~\ref{tab:const}.

The \textit{AM}Ag and \textit{AM}Cu molecules in the $X^1\Sigma^+$ electronic state have the largest permanent electric dipole moments ranging from 5.05$\,$debye for LiCu to 9.75$\,$debye for CsAg at the equilibrium distances and more for larger internuclear separations. To our best knowledge, these are one of the highest values predicted for neutral intermetallic dimers, comparable to \textit{AM}Au molecules~\cite{BelpassiJPCA06}. These values are also significantly larger than values for corresponding alkali-metal molecules, with the maximum value of 5.5$\,$debye for LiCs~\cite{AymarJCP05}.

The \textit{AM}Ag and \textit{AM}Cu molecules in the $a^3\Sigma^+$ electronic state have the smallest permanent electric dipole moments ranging from 0.017$\,$debye for LiCu to 0.13$\,$debye for CsAg at the equilibrium distances and a bit more for larger internuclear separations. These values are smaller or comparable to values for corresponding alkali-metal molecules~\cite{TomzaPRA13b}.

The \textit{AEM}Ag and \textit{AEM}Cu molecules in the $X^2\Sigma^+$ electronic state exhibit intermediate permanent electric dipole moments ranging from -0.81$\,$debye for BeCu to 5.08$\,$debye for RaAg at the equilibrium distances and more for larger internuclear separations. These values are similar or larger than values for corresponding alkali-metal--alkaline-earth-metal molecules~\cite{PototschnigPCCP16}.

The observed very large permanent electric dipole moments of the investigated ground-state molecules are directly related to the highly polarized covalent or even ionic nature of their chemical bonds, discussed in the previous subsection. The observed trends agree with the differences in atomic electronegativity. Permanent electric dipole moments are larger for the molecules based on the alkali-metal atoms than those based on the alkaline-earth-metal atoms. They are also slightly larger for the molecules based on the Ag atom than those based on the Cu atom. Finally, for all the molecules, they systematically increase with increasing the atomic number of involved alkali-metal or alkaline-earth-metal atoms, which correlates with decreasing their atomic electronegativity.  

Permanent electric dipole moments can also be used to measure the bond polarization and ionic character of the studied molecules~\cite{LiuPCCP20}. We quantify it by the ratio of the calculated permanent electric dipole moment of a given molecule, $d_e$, at the equilibrium distance, $R_e$, to the maximal possible value, $d_\text{max}=eR_e$, corresponding to a purely ionic molecule
\begin{equation}
\frac{d_{e}}{d_\text{max}}=\frac{d_{e}}{eR_{e}}\,.\\
\end{equation}
The calculated ratios are listed in Table~\ref{tab:const} and range from 0.0009 for the LiCu molecule in the $a^3\Sigma^+$ state to 0.628 for the CsAg molecule in the $X^1\Sigma^+$ state.

For the \textit{AM}Ag and \textit{AM}Cu molecules in the $X^1\Sigma^+$ electronic state, the ${d_{e}}/{d_\text{max}}$ ratio is about 0.46-0.63 implying the ionic character as large as 46$\,\%$-63$\,\%$, which is more than predicted by the differences of the atomic electronegativities in the previous subsection. Additionally, the permanent electric dipole moments for those molecules increase linearly with the interatomic distance in the vicinity of the interaction potential well, confirming their ionic character. The largest $d_e/d_\text{max}$ ratio for corresponding alkali-metal molecules is 0.32 for LiCs~\cite{AymarJCP05}.

For the \textit{AEM}Ag and \textit{AEM}Cu molecules in the $X^2\Sigma^+$ electronic state, the ${d_{e}}/{d_\text{max}}$ ratio implies the ionic character of about 20$\,\%$-35$\,\%$, except for the lightest alkaline-earth-metal atom, in agreement with the predictions based on the differences of the atomic electronegativities. The permanent electric dipole moments for those molecules also increase with the interatomic distance in the vicinity of the interaction potential well.

The observed increase of the permanent electric dipole moments with the interatomic distance is responsible for the unusual and significant increase of the permanent electric dipole moment with increasing the vibrational quantum number and decreasing the vibrational binding energy of the studied molecules. Permanent electric dipole moments of selected molecules in different vibrational levels of their ground electronic state as a function of the vibrational quantum number $v$ and binding energy $E_b$ are presented in Fig.~\ref{fig:dv}. They remain large even for highly excited vibrational levels, potentially allowing for new molecular control schemes. The largest values are 13.5$\,$debye for CsAg in the level with $v=170$ and $E_b=-879\,$cm$^{-1}$ and 12.1$\,$debye for CsCu with $v=116$ and $E_b=-3375\,$cm$^{-1}$ among the \textit{AM}Ag and \textit{AM}Cu molecules, and 6.0$\,$debye for BaAg with $v=87$ and $E_b=-2959\,$cm$^{-1}$ and 5.4$\,$debye for BaCu with $v=68$ and $E_b=-4070\,$cm$^{-1}$ among the \textit{AEM}Ag and \textit{AEM}Cu molecules. These extremely large permanent electric dipole moments combined with large reduced masses and small rotational constants, open the way for new quantum simulations of strongly interacting dipolar quantum many-body systems and controlled chemical reactions. 

The long-range dipolar interaction, 
\begin{equation}
E_{dd}(R,\theta)=\frac{d^2(1-3\cos^2\theta)}{R^3}\,,
\end{equation}
between the polarized CsAg molecules with the largest dipole moment of $d=13.5\,$debye will be as large as 28$\,$kHz at $R=1\,\mu$m or 220$\,$Hz at $R=5\,\mu$m. If molecules are not polarized by an external electric field, then in their ground rotational states, their interaction is dominated by  the effective isotropic term $-C_6^\text{rot}/R^6$, resulting from the dipolar interaction in the second-order of perturbation theory and given by the long-range coefficient 
\begin{equation}
C_6^\text{rot}=\frac{d^4}{6B_v}\,,
\end{equation}
where $B_v$ is the rotational constant for $v$ vibrational state. For the CsAg molecules, this coefficient exceeds $10^9$, which is two-to-three orders of magnitude larger than for alkali-metal dimers~\cite{ZuchowskiPRA13}.

For the completeness of the analysis, we also calculate the perpendicular $\alpha^\perp$ and parallel $\alpha^\parallel$ components of the static electric dipole polarizability tensor, which are important for the evaluation of intermolecular interactions and interactions with external electric or laser fields~\cite{DeiglmayrJCP08}. We report their values at the equilibrium distance, $\alpha_e^\perp$ and $\alpha_e^\parallel$, in Table~\ref{tab:const}. Interestingly, both components for the \textit{AM}Ag and \textit{AM}Cu molecules in the $X^1\Sigma^+$ electronic state are smaller than the asymptotic sum of atomic values, $\alpha_{AM}+\alpha_\text{Ag(Cu)}$, because the strong decrease of the atomic polarizability of $AM^+$ is not compensated by the increase of the atomic polarizability of Ag$^-$ or Cu$^-$, as compared to $AM$ and Ag or Cu, again in agreement with the ionic nature of those molecules. This effect is not pronounced in the \textit{AEM}Ag and \textit{AEM}Cu molecules, as expected for more covalent $AB$ metal dimers, where $\alpha^\perp<\alpha_{A}+\alpha_B$ and $\alpha^\parallel>\alpha_{A}+\alpha_B$~\cite{DeiglmayrJCP08}. The isotropic, $\bar{\alpha}=(2\alpha^\perp+\alpha^\parallel)/3$, and anisotropic, $\Delta\alpha=\alpha^\parallel-\alpha^\perp$, components can also be obtained from $\alpha^\perp$ and $\alpha^\parallel$.

\subsection{Chemical reactions}

The calculated potential well depths, $D_e$, and related dissociation energies, $D_0\approx D_e-\frac{1}{2}\omega_e$, may be used to assess the stability of the studied molecules against chemical reactions. In general, atom-exchange chemical reactions between ground-state heteronuclear molecules $AB$~\cite{ZuchowskiPRA10,TomzaPRA13b,SmialkowskiPRA20}
\begin{equation}
AB+AB \to A_2 + B_2
\end{equation}
are energetically possible if the sum of the dissociation energies of $A_2$ and $B_2$ products is larger or equal to the sum of the dissociation energies of reactants $AB$
\begin{equation}
D_0(A_2)+D_0(B_2)\ge 2\,D_0(AB)\,.
\end{equation}

Among the species investigated in this paper, the \textit{AM}Ag and \textit{AM}Cu molecules in the rovibrational ground state of the $X^1\Sigma^+$ electronic state are chemically stable against atom-exchange reactions for all alkali-metal atoms, e.g.,
\begin{equation}
2\,AM\text{Ag}(X^{1}\Sigma^+) \not\to \text{Ag}_2(X^{1}\Sigma^+) + AM_2(X^{1}\Sigma^+)\,.
\end{equation}

The AM\text{Ag} and AM\text{Cu} molecules in the weakly bound $a^3\Sigma^+$ electronic state, are reactive for all alkali-metal atoms, leading to Ag$_2$, Cu$_2$, and alkali-metal dimers in the $X^{1}\Sigma^+$ or $a^{1}\Sigma^+$ electronic state, e.g., 
\begin{equation}
\begin{split}
2\,AM\text{Ag}(a^{3}\Sigma^+) & \to \text{Ag}_2(X^{1}\Sigma^+) + AM_2(X^{1}\Sigma^+)\,,\\
2\,AM\text{Ag}(a^{3}\Sigma^+) & \to \text{Ag}_2(X^{1}\Sigma^+) + AM_2(a^{3}\Sigma^+)\,,\\
2\,AM\text{Ag}(a^{3}\Sigma^+) & \to \text{Ag}_2(a^{3}\Sigma^+) + AM_2(X^{1}\Sigma^+)\,,\\
2\,AM\text{Ag}(a^{3}\Sigma^+) & \to \text{Ag}_2(a^{3}\Sigma^+) + AM_2(a^{3}\Sigma^+)\,.
\end{split}
\end{equation}
Additionally, for those molecules, the spin relaxation reactions are possible
\begin{equation}
\begin{split}
2\,AM\text{Ag}(a^{3}\Sigma^+) & \to AM\text{Ag}(X^{1}\Sigma^+) + AM\text{Ag}(a^{3}\Sigma^+) \,,\\
2\,AM\text{Ag}(a^{3}\Sigma^+) & \to 2 AM\text{Ag}(X^{1}\Sigma^+) \,.
\end{split}
\end{equation}

The \textit{AEM}Ag and \textit{AEM}Cu molecules in the rovibrational ground state of the $X^2\Sigma^+$ electronic state are chemically stable against atom-exchange reactions for all alkaline-earth-metal atoms except MgAg and MgCu. For this two molecules, the following atom-exchange reaction is possible
\begin{equation}
2\,\text{MgAg}(X^{2}\Sigma^+) \to \text{Ag}_2(X^{1}\Sigma^+) + \text{Mg}_2(X^{1}\Sigma^+)\,.
\end{equation}

Except for the atom-exchange reactions, the trimers formation may be another path of chemical losses~\cite{ZuchowskiPRA10,TomzaPRA13b}
\begin{equation}
AB+AB \to A_2B + B\,,
\end{equation}
which is energetically possible if the dissociation energy of a $A_2B$ trimer product is larger or equal to the sum of the dissociation energies of reactants $AB$
\begin{equation}
D_0(A_2B)\ge 2\,D_0(AB)\,.
\end{equation}
However, three-body calculations for trimers containing Cu or Ag atoms are out of the scope of this paper.

The above-considered reactions, which are energetically forbidden in the lowest vibrational state ($v=0$), may be induced by the preparation or laser-field excitation of involved molecules to higher vibration levels.

\section{Summary and conclusions}
\label{sec:summary}

Ultracold gases of polar molecules, due to their rich and controllable internal molecular structure and intermolecular interactions, are excellent systems for experiments on precision measurements, quantum simulations of many-body physics, and controlled chemistry. Therefore, in this paper, we have proposed the formation and application of ultracold highly polar diatomic molecules containing a transition-metal copper or silver atom interacting with an alkali-metal or alkaline-earth-metal atom. To this end, we have employed state-of-the-art \textit{ab initio} electronic structure methods to study their ground-state properties in a comparative way. We have calculated potential energy curves, permanent electric dipole moments, spectroscopic constants, and leading long-range dispersion-interaction coefficients~\footnote{Full potential energy curves, permanent electric dipole moments, and electric dipole polarizabilities as a function of interatomic distance in a numerical form are available for all investigated molecules from the authors upon request.}.

We have predicted that the studied molecules in the ground electronic state are strongly bound with highly polarized covalent or ionic bonds resulting in significant permanent electric dipole moments, significantly larger than in alkali-metal molecules. We have found that maximal electric dipole moments, exceeding 13$\,$debye for CsCu and 6$\,$debye for BaAg, are for highly excited vibrational levels. To our best knowledge, these values are one of the highest predicted for neutral intermetallic molecules. We have also shown that most of the investigated molecules in the ground state are stable against atom-exchange chemical reactions.  The YbAg and YbCu molecules are expected to have proprieties similar to the considered \textit{AEM}Ag and \textit{AEM}Cu molecules due to similarities of the Yb atom to alkaline-earth-metal atoms.

The above peculiar properties of the studied highly polar molecules open the way for their application in ultracold physics and chemistry experiments. The extremely large permanent electric dipole moments combined with large reduced masses and small rotational constants for heavier molecules facilitate their orientation, alignment, and manipulation with external electric fields, on the one hand, and enhance intermolecular dipolar interactions, on the other hand. Thus, the studied molecules may be used in precision measurement of the electric dipole moment of the electron and the scalar-pseudoscalar interaction, as proposed for the RaCu and RaAg molecules~\cite{SunagaPRA19}. They may also be employed in quantum simulations of strongly interacting dipolar quantum many-body systems, where significant intermolecular interactions may be expected already at lower densities or between distant sites of an optical lattice or between optical tweezers. Finally, they may be exploited in quantum-controlled chemical reactions manipulated with external electric fields and vibrational excitations.

The investigated molecules can be formed in the same manner as the alkali-metal and alkali-metal--alkaline-earth-metal molecules, i.e., by using the magnetoassociation within the vicinity of the Feshbach resonance~\cite{KohlerRMP06} followed by the stimulated Raman adiabatic passage (STIRAP)~\cite{JonesRMP06}. A detailed analysis of their formation is out of the scope of this paper, but to facilitate their experimental realization and application, the excited molecular electronic states, photoassociation spectra, and specific laser-control schemes should be studied in the future.

\begin{acknowledgments}
We would like to thank Tatiana Korona for useful discussions. Financial support from the National Science Centre Poland (Grants No.~2015/19/D/ST4/02173 and No.~2016/23/B/ST4/03231) and the Foundation for Polish Science within the First Team program co-financed by the European Union under the European Regional Development Fund is gratefully acknowledged. The computational part of this research has been partially supported by the PL-Grid Infrastructure.
\end{acknowledgments}

\bibliography{CuAgX}

\end{document}